\documentstyle[aps,prl]{revtex}

\input psfig.sty

\newcommand{\stateone}{\protect{$\left | 1 \right>$}}
\newcommand{\statetwo}{\protect{$\left | 2 \right>$}}
\newcommand{\stateones}{\protect{$\left | 1 \right>\ $}}
\newcommand{\statetwos}{\protect{$\left | 2 \right>\ $}}

\draft

\begin{document}
\twocolumn[\hsize\textwidth\columnwidth\hsize\csname @twocolumnfalse\endcsname
\title{Watching dark solitons decay into vortex rings in a Bose-Einstein condensate}
\author{B.~P. Anderson,$^{1,2}$ P.~C. Haljan,$^{1}$ C.~A.~Regal,$^{3}$ D.~L.~Feder,$^{4}$ L.~A.~Collins,$^{5}$ C.~W.~Clark,$^{4}$ and E.~A.~Cornell$^{1,2}$}
\address{$^{1}$JILA, National Institute of Standards and Technology and Department of Physics, \\
University of Colorado, Boulder, Colorado 80309-0440}
\address{$^2$Quantum Physics Division, National Institute of
Standards and Technology, Boulder, CO 80305}
\address{$^3$Physics Department, Lawrence University, P.O.~Box 599, Appleton, WI 54912}
\address{$^4$Electron and Optical Physics Division, National Institute of
Standards and Technology, Gaithersburg, MD 20899-8410}
\address{$^5$Theoretical Division, Mail Stop B212, Los Alamos National
Laboratory, Los Alamos, NM 87545}
\date{\today}
\maketitle

\begin{abstract}

We have created spatial dark solitons in two-component
Bose-Einstein condensates in which the soliton exists in one of
the condensate components and the soliton nodal plane is filled
with the second component.  The filled solitons are stable for
hundreds of milliseconds.  The filling can be selectively
removed, making the soliton more susceptible to dynamical
instabilities. For a condensate in a spherically symmetric
potential, these instabilities cause the dark soliton to decay
into stable vortex rings.  We have imaged the resulting vortex
rings.\\

%PACS number(s): 03.75.Fi, 67.90.+z, 67.57.Fg, 32.80.Pj
\end{abstract}

%\pacs{03.75.Fi,67.90.+z,67.57.Fg,32.80.Pj}
]

%\narrowtext

Topological structures such as vortices and vortex rings have
fascinated scientists and mathematicians for centuries.  In a
quantum fluid, vortices have quantized flow around a
one-dimensional core where the density vanishes.  Vortex
\emph{lines}, which terminate at the boundaries of the quantum
fluid, have been observed in superfluid helium and
superconductors~\cite{donnelly,tilley}, and recently in
Bose-Einstein condensates (BECs)~\cite{matthews,madison}.
Similarly, vortex \emph{rings} are vortices whose cores are
closed loops; the poloidal quantized flow pattern resembles that
of a smoke ring.  Quantized vortex rings have been produced in
superfluid helium, and were first detected in pioneering work by
Rayfield and Reif~\cite{rayfield}. In this paper, we report the
first direct experimental observations of vortex rings in BECs.

Solitons are localized disturbances in a continuous nonlinear
medium that preserve their spatial profile due to a balance
between the effects of dispersion and
nonlinearity~\cite{solitonextra}. Dark solitons have been
previously created in single-component BECs using
phase-imprinting methods~\cite{darksolitonsinBEC}. These
structures are characterized by a local decrease in fluid
density (the depth). The macroscopic quantum phase of the BEC
differs on either side of the soliton; for a completely dark
(black) soliton, the depth is 100\% (a complete absence of
fluid), the phase offset is $\pi$, and the soliton velocity is
zero. We have created black solitons (having a fluid-free nodal
plane) in nearly spherically symmetric $^{87}$Rb BECs, and have
observed the subsequent decay of the solitons into vortex rings.

Unlike vortices and vortex rings, whose stability is ensured by
Kelvin's theorem~\cite{donnelly}, soliton stability depends on
the nonlinearity and geometry of the
medium~\cite{solitonstability}. Optical dark solitons in
self-defocusing nonlinear media have been observed to decay into
optical vortices via a `snake instability'
\cite{snakeinstability}, confirming predictions~\cite{jones}.   In general, dark
solitons in BECs are also expected to be inherently dynamically
unstable~\cite{josserand,solitonstabilityinBEC,carr,feder}.
Because both the velocity and depth of a soliton are determined
by its phase offset~\cite{reinhardt}, small local perturbations
of either the strength of the nonlinearity or of the depth give
the soliton a corresponding nonuniform transverse velocity
profile. Once the soliton begins to decay, it quickly breaks up
into more stable structures.   Within our spherically symmetric
BECs, the expected decay products are concentric vortex
rings~\cite{feder}. Because the theoretical background for
soliton decay in condensates has been previously described in
detail in Ref.~\cite{feder}, we limit our paper to primarily a
phenomenological presentation of data. Evidence for soliton
decay is obtained through direct images of the BEC density
distribution. We provide a visual comparison of data with
numerical simulations of soliton decay.

\emph{Experimental techniques} -- We create dark solitons
using concepts that we have previously applied to making singly
quantized vortices in BECs
\cite{matthews,williams}.  The magnetically trapped BEC can exist as a
superposition of two internal components, hyperfine states
\stateones $\equiv\left |F=1,m_{F}=-1
\right>$ and \statetwos $\equiv\left |F=2,m_{F}=1
\right>$.  Conversion between the two components is achieved with a
microwave field slightly detuned from the
\stateones to \statetwos internal conversion energy.
A ``modulation beam,'' an off-resonant laser beam whose small
spatial focus provides a requisite spatial selectivity, is
dithered rapidly across the BEC.  The laser beam induces a small
ac Stark shift in the \stateones to \statetwos transition
energy. The phase and amplitude of this effective fm modulation
varies from point-to-point in the sample in a spatial pattern
determined by the time-varying position of the optical
modulation beam; thus the microwave-induced inter-component
conversion varies in phase and amplitude across the BEC.

In our experiment, we start with a uniform-phase (i.e., ground
state) condensate of component
\statetwo.  To make a soliton, we choose a modulation beam pattern such that
inter-component conversion is suppressed in the middle of the
sample, whereas the upper and lower parts of the BEC undergo
conversion to component \stateones with an initial phase offset
of $\pi$ between the two parts. Uncontrolled variations in the
timing of the modulation cause the angular orientation of the
soliton nodal plane to vary randomly from one soliton to the
next.  For further details, see our earlier vortex work
\cite{matthews} and theory by Williams and Holland
\cite{williams}.  The soliton state contains about $3\times 10^5$
\stateones atoms, and the filling consists of about
$7\times 10^5$
\statetwos atoms.  The axial (vertical) trap frequency is 7.6(3)
Hz, the radial (horizontal) frequencies are 7.8(3) Hz, and the
condensates have a Thomas-Fermi (TF) radius of 28 $\mu$m. The
solitons are created at temperatures of $T=23(6)$ nK, or
$T/T_{c}=0.8(1)$, where $T_c$ is the BEC phase transition
temperature.

After removing the coupling field, we observe the trapped
excited-state condensates using nondestructive phase-contrast
imaging. Our probe laser is tuned such that only component
\stateones is visible, and these atoms appear bright on a dark
background. As described in \cite{matthews}, a filled-core
vortex appears as a dark hole in a bright atom cloud. Similarly,
a soliton with a filled nodal plane appears as a dark band that
divides the \stateones BEC.  This two-component BEC can be
described as a \emph{bright} soliton of component \statetwos
trapped within the \stateones dark soliton \cite{busch}.  The
filling material of
\statetwos atoms stabilizes the dark soliton against dynamical
instabilities; the filled solitons are observed to last for at
least 800 ms.

To study the dynamical instabilities of the solitons, we first
use resonant light pressure to selectively remove the \statetwos
atoms that fill the nodal plane.  The
\statetwos atoms are adiabatically removed over 100 ms, and a
bare dark soliton of component \stateones remains.  The soliton
node and any soliton decay products are then too small to be
observed while the condensate is held in the trap \cite{heal}.
We therefore remove the trapping potential and allow the
condensate to ballistically expand, causing the variations, or
`defects,' in the condensate density distribution to also expand
\cite{dalfovo}. These density defects are then resolvable. We obtain a final
near-resonance phase-contrast image of the expanded atom cloud,
using a probe detuning of 20 MHz.  Expansion imaging has
previously been used to detect bare vortex cores
\cite{madison,anderson}.

\emph{Numerical calculations} -- Before discussing experimental
data, we briefly describe results of numerical calculations. The
techniques employed have been described in detail in
Ref.~\onlinecite{feder}. The equilibrium configuration,
dynamics, and Bogoliubov excitation spectrum for a two-component
condensate at low temperatures may all be obtained from the
appropriate three-dimensional time-dependent Gross-Pitaevskii
(GP) equation~\cite{pu}, with $N_1=3\times 10^5$ \stateones
atoms in a state initially constrained to have odd parity along
one axis, and $0\leq N_2\leq 7\times 10^5$ \statetwos atoms in
an even-parity state. The spherical trap frequency is chosen to
be $\nu_0=7.8$~Hz, and the intra- and inter-species scattering
lengths are
$(a_{11}\,,a_{22}\,,a_{12})=5.5(1.03\,,0.97\,,1.0)$~nm,
respectively~\cite{hall}. The behavior of the BEC in full
two-component simulations, in which we approximate a slow
removal of the
\statetwos atoms, is found
\begin{figure}  % FILE: fig1_bpa.eps
\psfig{figure=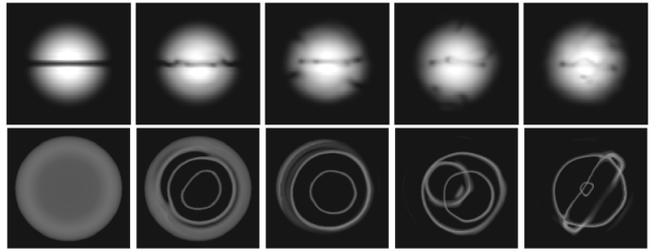,width=1\linewidth,clip=}
\caption{Results of numerical simulations showing the decay of a black
soliton in a BEC.  The simulation corresponds to $3\times 10^5$
$^{87}$Rb atoms in a spherically symmetric trap with frequency
7.8~Hz.  Successive frames are shown at $50$~ms intervals, with
the first frame at $100$~ms after the start of the simulation.
The first row shows the density profile of the condensate,
integrated down an axis parallel to the soliton plane. The
low-density regions within the cloud are also rendered (second
row), with views perpendicular to the soliton plane.}
\label{simulate}
\end{figure}
\noindent to be qualitatively similar to
simulations where we set $N_2=0$ at the outset.  In all cases,
the soliton is found to undergo a snake instability, decaying
into vortex rings.

Indeed, the Bogoliubov spectrum contains modes with complex
frequencies for all $N_2$ considered; such modes have been shown
to drive the soliton instability~\cite{feder}. For $N_2\gtrsim
4\times 10^5$, only one such mode remains, with imaginary
frequency of magnitude $\nu\sim(4\times 10^4/N_2)\nu_0$.
Assuming a soliton decay time $\propto\nu^{-1}$, a filled
soliton with $N_2=7\times 10^5$ is expected to be stable for
longer than 2~s, consistent with experimental observations.

The results of a one-component ($N_2=0$) simulation are shown in
Fig.~\ref{simulate}. The soliton decays into three nearly
concentric vortex rings approximately 130 ms after initial
formation. In general, the rings tend to migrate towards the
condensate surface, and may then grow, shrink, and reconnect
with each other.  The innermost ring is the most stable,
remaining intact for at least 150~ms in the absence of a
collision with another ring.  Along an imaging axis parallel to
the plane of a given vortex ring, the integrated column density
reveals two dips in the density distribution, connected by a
fainter line. The ring, however, is not constrained to lie in
any given plane, and may bend and tilt, producing other types of
images.  In elongated traps with aspect ratios of two or higher,
simulations indicate that a soliton can decay directly into
nearly parallel vortex lines rather than rings; however, this
behavior is not obtained for spherical traps.

\emph{Experimental results} -- In our experimental cycle, we first create
a filled dark soliton and obtain a non-destructive image of the
initial trapped (filled) soliton. Over the next 100 ms, we
remove the
\statetwos atoms from the soliton nodal plane until only the \stateones atoms remain.
The trapped single-component condensate is then held for a
variable hold time before being allowed to ballistically expand
for 56 ms
\cite{expansiontechnique}. Finally, the expanded density distribution is recorded.

Images of the decayed solitons typically reveal two dips
\begin{figure}  % FILE: fig2_bpa.eps
\psfig{figure=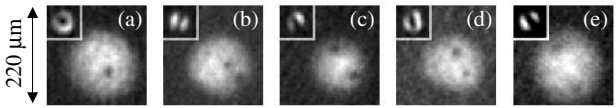,width=1\linewidth,clip=}
\caption{Typical images of expanded condensates, and their initial states
(insets) before the \statetwos atoms were removed. All images
are shown at the indicated scale. (a) A vortex state, (b)-(e)
the decay products of solitons. Image (e) was taken with a hold
time of 500 ms; all other images were taken with a hold time of
0 ms, in addition to the 100 ms
\statetwos removal and the 56 ms expansion [22].} \label{expanded1}
\end{figure}
\noindent in the condensate density distribution, as shown in
Fig.~\ref{expanded1}(b)-(e).  While our images are consistent
with the expected signatures of vortex rings, they are
inconclusive evidence for the decay of solitons into vortex
rings; from a single image containing only a pair of density
dips, we can not differentiate between a
\emph{single vortex ring} and a
\emph{pair of vortex lines}.  For comparison,
an expanded vortex line is shown in Fig.~\ref{expanded1}(a). We
believe that our signal-to-noise levels may limit our ability to
see a dark line connecting the two clear density defects that
would indicate the sides of a ring. We also note that
high-frequency excitations visible in $T=0$ simulations would be
quickly damped at finite temperatures, and thus not seen in the
experiment.

\begin{figure}  % FILE: fig3_bpa.eps
\psfig{figure=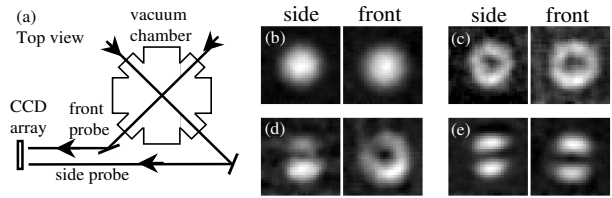,width=1\linewidth,clip=} \caption{(a) An
illustration of our probe beam imaging paths.  (b)-(e)
Simultaneous 100-$\mu$m-square images of trapped condensates
from the front and side imaging directions. Shown are (b) a
ground state condensate of component
\stateone, (c) a shell of \stateones atoms around a central ball
of \statetwos atoms [21] , (d) a \stateones vortex filled with
\statetwos atoms, and (e) a horizontally oriented dark soliton of
component \stateone, filled with \statetwos atoms.}
\label{beampathfig}
\end{figure}

Because detection of two-dimensional topological structures,
such as dark solitons and vortex rings, can be enhanced by
probing along two orthogonal directions, we added an additional
probe beam.  In our improved apparatus, the second probe beam is
orthogonal to the original beam, and propagates horizontally (as
does the original). The beams intersect at the condensate.
Phase-contrast imaging is used independently with each beam
path, and the probe beams propagate towards different sides of a
single charge-coupled device (CCD) camera array. The apparatus
allows us to take simultaneous pictures of condensates from the
`front' (original) and the `side' (added) directions using a
single camera, as illustrated in Fig.~\ref{beampathfig}(a).
While we can observe the full depth of the filled soliton node
along the front direction, regardless of the initial
orientation, the nodal plane is usually at an oblique angle to
the side probe, and soliton contrast is thus reduced. Examples
of simultaneous images of trapped condensates from the front and
side are shown in Figs.~\ref{beampathfig}(b)-(e). In the pair of
images shown in Fig.~\ref{beampathfig}(d), the filled core of a
vortex is visible as the dark hole in the front image, and as
the dark band in the side image.  In Fig.~\ref{beampathfig}(e),
a filled horizontally oriented dark soliton appears as a band
across the condensate in each probing direction, demonstrating
that the soliton splits the BEC into two sections.

\begin{figure}  % FILE: fig4_bpa.eps
\psfig{figure=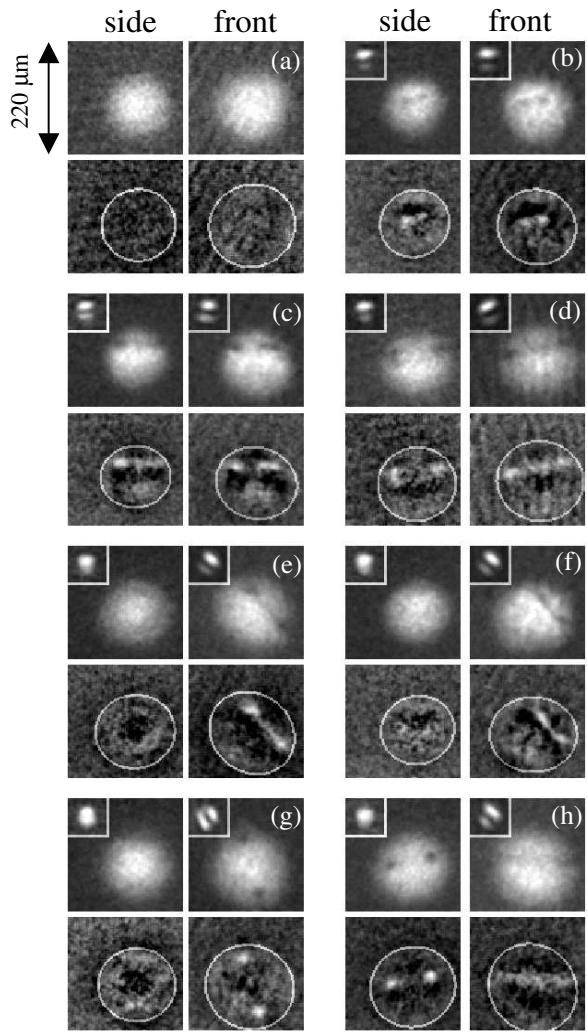,width=1\linewidth,clip=} \caption{Images
of expanded condensates and their initial states (insets), as
viewed along both imaging axes.  Each subfigure consists of two
expansion images (top row of each subfigure) and two images of
residuals plots (bottom row), obtained after subtracting a TF
fit of each expansion image.  The white regions in the residuals
correspond to depletion of fluid in the BEC. The residuals are
superimposed with white ellipses outlining the fit TF profiles
of the condensates, for position reference. (a) An expanded
ground state condensate. (b)-(h) The decay products of solitons.
The hold times (between
\statetwo-atom removal and BEC expansion) were 0 ms for images
(a)-(c),(f), and (g); 50 ms for (d) and (e); and 150 ms for
(h).}\label{expanded2}
\end{figure}

The use of two orthogonal probe beams confirms that dark
solitons indeed decay into vortex rings.  With the two beam
paths, we have observed pairs of density dips in simultaneous
side and front expanded images of decayed
\noindent  solitons.  In each
image set shown in Figs.~\ref{expanded2}(b)-(d), pairs of
density dips lie nearly horizontally and at corresponding
locations within the two BEC images.  We have also obtained
images in which a faint line corresponding to a decrease in
integrated column density can be observed between two defects in
one of the expanded images, as shown in Figs.~\ref{expanded2}(e)
and (f), providing further striking demonstrations of soliton
decay into vortex rings. Figures~\ref{expanded2}(e)-(g)
emphasize that when the vortex ring is not close to horizontal,
clear pairs of density dips are not seen in the side images.
However, a weak ring-shaped halo may be discerned in the side
image residuals of Figs.~\ref{expanded2}(e) and (g), when the
ring is in a near-vertical plane orthogonal to the side probe
direction. Although the halo signal is only slightly above the
noise of the residuals, it can be noticed when compared to the
residuals for an expanded ground state condensate, shown in
Fig.~\ref{expanded2}(a).  Images in Figs.~\ref{expanded2}(b)
through (f) represent the best 10\% of our data showing solitons
decaying into rings; Fig.~\ref{expanded2}(g) is typical.

Numerical as well as experimental data show great variability in
the types of images that can be expected after a soliton decays.
For example, Fig.~\ref{expanded2}(h) is a rare example showing
two clear density dips in the side image, but here a clear line
corresponding to the defect extends through the entire
condensate in the front image and residuals.  It is possible
that this image shows soliton decay into two vortex lines rather
than a ring, or breakup of a ring into vortex lines. We defer to
a future paper a discussion of the diversity of our images,
which may result from interesting vortex ring dynamics,
including vortex ring bending, tilting, and reconnections.

Small perturbations unavoidably induce a dark soliton to decay.
In a spherically symmetric BEC, a soliton decays into a
relatively stable vortex ring.  In addition to demonstrating
this nonlinear decay process, our experiment shows that
condensates can indeed support vortex rings, which may be
detected via expansion imaging. Future experiments may include
studies of vortex ring stability, lifetime, and dynamics
\cite{jackson}, as well as investigations into vortex structures
created by impurity motion above a superfluid critical velocity
\cite{onofrio}.
\noindent  Possibilities to create vortex rings by other
means may also be explored, such as by passing objects through
the condensate
\cite{jackson2}.

We thank Seamus Davis, Murray Holland, and Carl Wieman for
helpful discussions. This work was supported by funding from
NSF, ONR, and NIST.  L.A.C. acknowledges funding from the DOE.


\begin{thebibliography}{88}
\bibitem{heal} The spatial width of a bare dark soliton or vortex
core in a pre-expanded sample is on the order of the BEC healing
length, about 0.7 $\mu$m for our condensates.
\bibitem{expansiontechnique} Condensates expand slowly when released from our weak trap.
We get larger final spatial distributions by giving the atoms a
6-ms preliminary ``squeeze'' at high spring constants, followed
by a 50-ms expansion period.  Dynamical soliton evolution and
decay begins during the \statetwos removal, and continues
through about the first few milliseconds of expansion;
thereafter, the decay pattern is essentially frozen into the
expanding sample.
\end{thebibliography}
\end{document}